\documentstyle[nato,graphicx]{crckapb}

\def\nm{\hbox{$\nu_\mu$ }}
\def\nt{\hbox{$\nu_\tau$ }}
\def\rp{$R_p \hspace{-1em}/\;\:$ }
\def\beq{\begin{equation}}
\def\eeq{\end{equation}}
\def\Slash#1{#1\!\!\!\! /}
\def\slash#1{#1\!\!\! /}
\def\beqa{\begin{eqnarray}}
\def\eeqa{\end{eqnarray}}
\def\vev#1{\left\langle #1\right\rangle}

\def\ra{\rightarrow}

\def\vb#1{\vbox to #1 pt{}}
\def\ifmath#1{\relax\ifmmode #1\else $#1$\fi}
\def\half{\ifmath{{\textstyle{1 \over 2}}}}
\def\quarter{\ifmath{{\textstyle{1 \over 4}}}}
\def\nn{\nonumber}
\def\21{$SU(2) \otimes U(1)$}
\def\beqa{\begin{eqnarray}}
\def\eeqa{\end{eqnarray}}
\def\ds{\displaystyle}
\def\bold#1{\setbox0=\hbox{$#1$} 
     \kern-.025em\copy0\kern-\wd0 
     \kern.05em\copy0\kern-\wd0 
     \kern-.025em\raise.0433em\box0 }
\def\ni{\noindent}
\def\Table#1{Table~(\ref{#1})}
\def\Eq#1{Eq. (\ref{#1})}

\def\Fig#1{Fig.~(\ref{#1})}

\newcommand {\ignore}[1]{}

\begin{opening}
\title{SOLVING THE SOLAR AND ATMOSPHERIC NEUTRINO\protect\\ 
       PROBLEMS WITH SUPERSYMMETRY}

\subtitle{Talk given at the NATO Advanced 
Study Institute 2000, Cascais, Portugal, 26 June - 7 July, 2000.}

\author{J.C. ROMAO}

\institute{Instituto Superior T\'ecnico, Departamento de F\'{\i}sica\\
           Av. Rovisco Pais, 1, 1049-001 Lisboa, Portugal}

\end{opening}

\runningtitle{A SUPERSYMMETRIC SOLUTION ...}

\begin{document}

\begin{abstract}
The simplest unified extension of the Minimal Supersymmetric Standard
Model with bi-linear R--Parity violation provides a predictive scheme
for neutrino masses which can account for the observed atmospheric and
solar neutrino anomalies.
Despite the smallness of neutrino masses R-parity violation is
observable at present and future high-energy colliders, providing an
unambiguous cross-check of the model.
\end{abstract}

\section{Introduction}

The recent announcement of high statistics atmospheric neutrino data
by the SuperKamiokande collaboration \cite{Fukuda:1998mi} has
confirmed the deficit of muon neutrinos, especially at small zenith
angles, opening a new era in neutrino physics. 
Although there may be alternative solutions of the atmospheric
neutrino anomaly ~\cite{Gonzalez-Garcia:1998hj} it is fair to say that
the simplest interpretation of the data is in terms of \nm to \nt
flavour oscillations with maximal mixing. This excludes a large mixing
among $\nu_{\tau}$ and $\nu_e$~\cite{Fukuda:1998mi}, in agreement also
with the Chooz reactor data.   On the other hand the
persistent disagreement between solar neutrino data and theoretical
expectations~\cite{BP98} has been a long-standing problem in physics. 
Recent solar neutrino
data~\cite{Smy:1999tt} are consistent with both
vacuum oscillations and MSW conversions. In the latter case one can
have either the large or the small mixing angle solutions, with a
slight trend towards the latter~\cite{MSW99}. 

Many attempts have appeared in the literature to explain the
data. Here we review recent  results~\cite{numass} obtained in a 
model~\cite{epsmodel} which is a simple
extension of the MSSM with with bilinear R-parity violation (BRPV). 
This model, despite being a minimal extension of the MSSM, can explain
the solar and atmospheric neutrino data. Its most attractive feature
is that it gives definite predictions for accelerator physics for the
same range of parameters that explain the neutrino data.

\section{Broken R--parity}
In the past most discussions of supersymmetric (SUSY)
phenomenology assumed R--parity ($R_P$) conservation where,
\beq
R_P=(-1)^{2J +3B +L}
\eeq
This implies that SUSY particles are pair produced, 
every SUSY particle decays into another SUSY particle and that
there is a {\it LSP} that it is stable.
But this is just an {\it ad hoc} assumption without a deep
justification. 
In this talk we will review how $R_P$ can be broken, either
spontaneously or explicitly, and discuss the most important features
of these models \cite{brasil98}.

\subsection{Spontaneously Broken R-Parity}

\subsubsection{The Original Proposal}

In the original proposal~\cite{OriginalSBRP} the content was just the 
MSSM and the breaking was induced by
\beq
\vev{\tilde{\nu}_{\tau}} = v_L
\eeq
The problem with this model was that the Majoron $J$ coupled to $Z^0$ with 
gauge strength and therefore the decay
$Z^0 \rightarrow \rho_L J$ contributed to the invisible $Z$ width the
equivalent of half a (light) neutrino family. After LEP I this was
excluded.

\subsubsection{A Viable Model for SBRP}

The way to avoid the previous difficulty 
is to enlarge the model and make $J$ mostly out of {\it
isosinglets}. This was proposed by Masiero and Valle~\cite{SBRpV}. The
content is the MSSM plus a few Isosinglet Superfields that carry
lepton number.
The model is defined by the superpotential \cite{SBRpV},
\beqa
W&=&\phantom{+}h_u u^c Q H_u + h_d d^c Q H_d + h_e e^c L H_d \nn \\ 
&&+h_0 H_u H_d \Phi + \frac{\lambda}{3!}\Phi^3 
+ h_{\nu} \nu^c L H_u + h \Phi \nu^c S 
\eeqa
where the lepton number assignments are shown in \Table{table0}.
\begin{table}
\begin{center}
\caption{Lepton number assignments.}
\begin{tabular}{lccccc} 
\hline
Field & $L$ & $e^c$ & $\nu^c$ & $S$ & others  \\ 
Lepton \# & $1$ & $-1$ & $-1$& $1$ & $0$ \\ 
\hline
\label{table0}
\end{tabular}
\end{center}
\end{table}
The spontaneous breaking of R parity
and lepton number is driven by \cite{SBRpV}
\beq
v_R = \vev {\tilde{\nu}_{R\tau}} \quad
v_S = \vev {\tilde{S}_{\tau}} \quad
v_L = \vev {\tilde{\nu}_{\tau}}
\eeq
The electroweak breaking and fermion masses arise from
\beq
\vev {H_u} = v_u ~~~~~
\vev {H_d} = v_d
\eeq
with $v^2 = v_u^2 + v_d^2$ fixed by the W mass.
The Majoron is given by the imaginary part of 
\beq
\frac{v_L^2}{V v^2} (v_u H_u - v_d H_d) +
              \frac{v_L}{V} \tilde{\nu_{\tau}} -
              \frac{v_R}{V} \tilde{{\nu^c}_{\tau}} +
              \frac{v_S}{V} \tilde{S_{\tau}}
\eeq
where $V = \sqrt{v_R^2 + v_S^2}$. 
Since the Majoron
is mainly an \21 singlet it does not contribute to the
invisible $Z^0$ decay width.

\subsubsection{Some Results on SBRP}

The SBRP model has been extensively studied. The implications for
accelerator and non--accelerator
physics have been presented  before and we will not discuss them here
\cite{brasil98}. In this talk we will only review 
the results for neutrinos. 
Neutrinos are massless at Lagrangian level but get mass from the
mixing with neutralinos\cite{paulo+npb}. 
Neutrinos mix and the mixing is related to the the 
coupling matrix $h_{\nu_{ij}}$. This matrix  has to be non diagonal in
generation space to allow
\beq
\nu_{\tau} \rightarrow \nu_{\mu} + J 
\eeq
and therefore evading~\cite{paulo+npb} the {\it Critical Density Argument} against
$\nu's$ in the MeV range. 
In the {\it SM} BBN arguments~\cite{bbnothers} rule out $\nu_{\tau}$ 
masses in the range
\beq
0.5\ MeV < m_{\nu_{\tau}} < 35 MeV
\eeq
We have shown~\cite{bbnpaper} that {\it SBRP} models can evade that 
constraint due to new annihilation channels
\beq
\nu_{\tau} \nu_{\tau} \rightarrow J J 
\eeq

\subsection{Explicitly Broken R-Parity}

The most general superpotential $W$ with the particle content
of the MSSM is given by \cite{epsmodel}

\beq
W= W_{MSSM} + W_{\Slash{R}}
\eeq
where 
\begin{equation}
W_{MSSM}=\varepsilon_{ab}\left[
 h_U^{ij}\widehat Q_i^a\widehat U_j\widehat H_u^b
+h_D^{ij}\widehat Q_i^b\widehat D_j\widehat H_d^a
+h_E^{ij}\widehat L_i^b\widehat R_j\widehat H_d^a 
 -\mu\widehat H_d^a\widehat H_u^b \right]
\label{WMSSM}
\end{equation}
and 
\begin{equation}
W_{\Slash{R}}=\varepsilon_{ab}\left[
 \lambda_{ijk}\widehat L_i^a\widehat L_j^b\widehat R_k
+\lambda^{'}_{ijk}\widehat D_i \widehat L_j^a\widehat Q_k^b 
+\lambda^{''}_{ijk} \widehat D_i \widehat D_j\widehat U_k 
+ \varepsilon_{ab}\, 
\epsilon_i\widehat L_i^a\widehat H_u^b \right]
\label{WRPV}
\end{equation}
where $i,j=1,2,3$ are generation indices, $a,b=1,2$ are $SU(2)$
indices. To these we also have to add the soft supersymmetry breaking
terms~\cite{numass}.

\ignore{
\noindent
The set of soft supersymmetry
breaking terms are
\beq
V^{soft}=V^{soft}_{ MSSM}+V^{soft}_{\Slash{R}}
\eeq
\begin{eqnarray}
V^{soft}_{\hbox{\tiny MSSM}}&\hskip -3mm=\hskip -3mm&
M_Q^{ij2}\widetilde Q^{a*}_i\widetilde Q^a_j+M_U^{ij2}
\widetilde U^*_i\widetilde U_j+M_D^{ij2}\widetilde D^*_i 
\widetilde D_j
+M_L^{ij2}\widetilde L^{a*}_i\widetilde L^a_j
+M_R^{ij2}\widetilde R^*_i\widetilde R_j \nn \\
&\hskip -3mm\hskip -3mm&
+m_{H_d}^2 H^{a*}_d H^a_d+m_{H_u}^2 H^{a*}_u H^a_u 
- \left[\half \sum_{i=1}^{3} M_i\lambda_i\lambda_i
+\varepsilon_{ab}\left(
A_U^{ij}\widetilde Q^a_i\widetilde U_j H_u^b \right. \right. \nn \\
&\hskip -3mm\hskip -3mm&
\left. \vb{18.5} \left.
+A_D^{ij}\widetilde Q^b_i\widetilde D_j H_d^a 
+A_E^{ij}\widetilde L^b_i\widetilde R_j H_d^a 
-B\mu H_d^a H_u^b \right) +h.c. \right]
\, ,
\label{SoftMSSM}
\end{eqnarray}
and
\beq
V^{soft}_{\Slash{R}}=
\varepsilon_{ab}\left[
A_{\lambda}^{ij}\widetilde L^a_i\widetilde L_j^b \widetilde R_k
+A_{\lambda'}^{ijk}\widetilde D_i\widetilde L_j^a \widetilde Q_k^b 
+A_{\lambda^{''}}^{ij}\widetilde D_i\widetilde D_j \widetilde U_k 
+B_i\epsilon_i\widetilde L^a_i H_u^b \right] + h.c.
\label{SoftRPV}
\eeq
}

\section{Bilinear R-Parity Violation (BRPV)}
\subsection{The Model}

The superpotential $W$ is
given by 
\beq
W
\hskip -1mm=\hskip -0.5mm
\varepsilon_{ab}\! \left[
 h_U^{ij}\widehat Q_i^a\widehat U_j\widehat H_u^b
\!+h_D^{ij}\widehat Q_i^b\widehat D_j\widehat H_d^a
\!+h_E^{ij}\widehat L_i^b\widehat R_j\widehat H_d^a 
\! -\mu\widehat H_d^a\widehat H_u^b
\!+\epsilon_i\widehat L_i^a\widehat H_u^b\right]
\eeq
while the set of soft supersymmetry
breaking terms are
\begin{eqnarray}
V_{soft}&\hskip -5mm=\hskip -5mm&
M_Q^{ij2}\widetilde Q^{a*}_i\widetilde Q^a_j+M_U^{ij2}
\widetilde U^*_i\widetilde U_j+M_D^{ij2}\widetilde D^*_i
\widetilde D_j+M_L^{ij2}\widetilde L^{a*}_i\widetilde L^a_j
+M_R^{ij2}\widetilde R^*_i\widetilde R_j
\nn \\
&&\hskip -5mm
+m_{H_d}^2 H^{a*}_d H^a_d+m_{H_u}^2 H^{a*}_u H^a_u 
- \left[\sum_i \half M_i\lambda_i\lambda_i
+\varepsilon_{ab}\left(
A_U^{ij}\widetilde Q^a_i\widetilde U_j H_u^b \right. \right. \nn \\
&&\hskip -5mm
\left. \vb{18} \left.
+A_D^{ij}\widetilde Q^b_i\widetilde D_j H_d^a
+A_E^{ij}\widetilde L^b_i\widetilde R_j H_d^a 
-B\mu H_d^a H_u^b+B_i\epsilon_i\widetilde L^a_i H_u^b \right) 
+ h.c \right]
\end{eqnarray}

\ni
The bilinear
R-parity violating term {\sl cannot} be eliminated by superfield
redefinition.
The reason is \cite{marco} 
that the bottom Yukawa coupling, usually neglected,
plays a crucial role in splitting
the soft-breaking parameters $B$ and $B_i$ as well as the scalar
masses $m_{H_d}^2$ and $M_L^{2}$, assumed to be equal at the
unification scale.

\ignore{
\ni
The electroweak symmetry is broken when the VEVS of 
the two Higgs doublets $H_d$
and $H_u$, and the sneutrinos.
\beq
\vev{H_d}={{{1\over{\sqrt{2}}}\, v_d}\choose{0}}\,,
\vev{H_u}={{0}\choose{{1\over{\sqrt{2}}}\, v_u }}\,,
\vev{\widetilde L_i}={{{1\over{\sqrt{2}}}
v_i}\choose{0}}
\eeq
The gauge bosons $W$ and $Z$ acquire masses
\beq
m_W^2=\quarter g^2v^2 \quad ; \quad m_Z^2=\quarter(g^2+g'^2)v^2
\eeq
where
\beq
v^2\equiv v_d^2+v_u^2+v_1^2+v_2^2+v_3^2=(246 \; {\rm GeV})^2
\eeq
The full scalar potential may be written as
\beq
V_{total}  = \sum_i \left| { \partial W \over \partial z_i} \right|^2
	+ V_D + V_{soft} + V_{RC}
\eeq
where $z_i$ denotes any one of the scalar fields in the
theory, $V_D$ are the usual $D$-terms, $V_{soft}$ the SUSY soft
breaking terms, and $V_{RC}$ are the 
one-loop radiative corrections. 

\ni
In writing $V_{RC}$ we  use the diagrammatic method and find 
the minimization conditions by correcting to one--loop the tadpole
equations. 
This method has advantages with respect to the effective potential when
we calculate the one--loop corrected scalar masses.
The scalar potential contains linear terms
\beq
V_{linear}=t_d\sigma^0_d+t_u\sigma^0_u+t_i\tilde\nu^R_{i}
\equiv t_{\alpha}\sigma^0_{\alpha}\,,
\eeq
where we have introduced the notation
\beq
\sigma^0_{\alpha}=(\sigma^0_d,\sigma^0_u,\nu^R_1,\nu^R_2,\nu^R_3)
\eeq
and $\alpha=d,u,1,2,3$. The one loop tadpoles are
\begin{eqnarray}
t_{\alpha}&=&t^0_{\alpha} -\delta t^{\overline{MS}}_{\alpha}
+T_{\alpha}(Q)\cr
\vb{22}
&=&t^0_{\alpha} +T^{\overline{MS}} _{\alpha}(Q)\nn
\end{eqnarray}
where $T^{\overline{MS}} _{\alpha}(Q)\equiv -\delta t^{\overline{MS}}_{\alpha}
+T_{\alpha}(Q)$ are the finite one--loop tadpoles.

\subsection{Main Features}
}

\ni
The BRPV model is a 1(3) parameter(s) generalization of the MSSM.
It can be thought as an {\bf effective} model
showing the more important features of the SBRP--model at the weak
scale. 
The mass matrices, charged and neutral currents, are similar to the
SBRP--model if we identify
\beq
\epsilon \equiv v_R h_{\nu}
\eeq
The model has the MSSM as a limit when $\epsilon_i\ra 0$.

\subsection{Radiative Breaking}

At $Q = M_{GUT}$ we assume the standard minimal supergravity
unifications assumptions, 
\beqa
&&A_t = A_b = A_{\tau} \equiv A \:, B=B_2=A-1 \:, \nn \\
&&
m_{H_d}^2 = m_{H_u}^2 = M_{L}^2 = M_{R}^2 = 
M_{Q}^2 =M_{U}^2 = M_{D}^2 = m_0^2 \:, \nn \\
&&
M_3 = M_2 = M_1 = M_{1/2} 
\eeqa
In order to determine the values of the Yukawa couplings and of the
soft breaking scalar masses at low energies we first run the RGE's from
the unification scale $M_{GUT} \sim 10^{16}$ GeV down to the weak
scale. For details see~\cite{numass,epsmodel}.

\ignore{

We randomly give values at the unification scale
for the parameters of the theory. 
\ignore{
\beq
\begin{array}{ccccc}
10^{-2} & \leq &{h^2_t}_{GUT} / 4\pi & \leq&1 \cr
10^{-5} & \leq &{h^2_b}_{GUT} / 4\pi & \leq&1 \cr
-3&\leq&A/m_0&\leq&3 \cr
0&\leq&\mu^2_{GUT}/m_0^2&\leq&10 \cr
0&\leq&M_{1/2}/m_0&\leq&5 \cr
10^{-2} &\leq& {\epsilon^2_i}_{GUT}/m_0^2 &\leq& 10\cr 
\end{array}
\eeq
The values of $h_{e}^{GUT},h_{\mu}^{GUT},h_{\tau}^{GUT}$ are 
defined in such a way
that we get the charged lepton  masses correctly. 
As the charginos mix with the 
leptons some care is necessary to accomplish this~\cite{numass}.
}
After running the RGE we have a 
complete set of parameters, Yukawa couplings and soft-breaking masses 
$m^2_i(RGE)$ to study the minimization. 
This is done by the following method: we solve the minimization
equations for the soft masses squared. This is easy because those
equations are linear on the soft masses squared. The values obtained
in this way, that we call $m^2_i$ are not equal to the values
$m^2_i(RGE)$ that we got via RGE. To achieve equality we define a function
\beq
\eta= max \left(  \frac{\ds m^2_i}{\ds m^2_i(RGE)},
\frac{\ds m^2_i(RGE)}{\ds m^2_i}
\right) \quad \forall i 
\eeq
with the obvious property that
\beq
\eta \ge 1
\eeq
Then we adjust the parameters to minimize $\eta$.
}

\section{Tree Level Neutrino Masses and Mixings}

\subsection{Neutral fermion mass matrix}

In the basis $\psi^{0T}= 
(-i\lambda',-i\lambda^3,\widetilde{H}_d^1,\widetilde{H}_u^2,
\nu_{e}, \nu_{\mu}, \nu_{\tau} )$ 
the neutral fer\-mions mass terms in the Lagrangian are given by 
\beq
{\cal L}_m=-\frac 12(\psi^0)^T{\bold M}_N\psi^0+h.c.   
\eeq
where the neutralino/neutrino mass matrix is 
\beq
{\bold M}_N=\left[  
\begin{array}{cc}  
{\cal M}_{\chi^0}& m^T \cr
m & 0 \cr
\end{array}
\right]
\eeq
with
\beq
{\cal M}_{\chi^0}=\left[  
\begin{array}{cccc}  
M_1 & 0 & -\frac 12g^{\prime }v_d & \frac 12g^{\prime }v_u \cr
0 & M_2 & \frac 12gv_d & -\frac 12gv_u \cr
-\frac 12g^{\prime }v_d & \frac 12gv_d & 0 & -\mu  \cr
\frac 12g^{\prime }v_u & -\frac 12gv_u &  -\mu & 0  \cr
\end{array}  
\right] 
\quad ; \quad
m=\left[  
\begin{array}{c}
a_1 \cr
a_2 \cr
a_3 
\end{array}  
\right] 
\eeq
where $a_i=(-\frac 12g^{\prime }v_i, \frac 12gv_i, 0,\epsilon_i)$. 
This neutralino/neutrino mass matrix is diagonalized by 
\beq
{\cal  N}^*{\bold M}_N{\cal N}^{-1}={\rm diag}(m_{\chi^0_1},m_{\chi^0_2}, 
m_{\chi^0_3},m_{\chi^0_4},m_{\nu_1},m_{\nu_2},m_{\nu_3}) 
\label{eq:NeuMdiag} 
\eeq

\subsection{Approximate diagonalization of mass matrices}

If the \rp parameters are small, then
\beq
\xi = m \cdot {\cal M}_{\chi^0}^{-1}\hskip 10mm
\Rightarrow\hskip 10mm \forall \xi_{ij} \ll 1
\eeq
one can find an approximate solution for the mixing  matrix ${\cal N}$.
Explicit solutions can be found in Ref.~\cite{numass}.
\ignore{
Explicitly
\begin{eqnarray}
\xi_{i1} &=& \frac{g' M_2 \mu}{2 det({\cal M}_{\chi^0})}\Lambda_i \nn \\
\xi_{i2} &=& -\frac{g M_1 \mu}{2 det({\cal M}_{\chi^0})}\Lambda_i \nn \\
\xi_{i3} &=& - \frac{\epsilon_i}{\mu} + 
          \frac{(g^2 M_1 + {g'}^2 M_2) v_2}
               {4 det({\cal M}_{\chi^0})}\Lambda_i \nn \\
\xi_{i4} &=& - \frac{(g^2 M_1 + {g'}^2 M_2) v_1}
               {4 det({\cal M}_{\chi^0})}\Lambda_i
\end{eqnarray}
where
\beq
\Lambda_i = \mu v_i + v_d \epsilon_i
\eeq
}
In leading order in $\xi$ the mixing matrix ${\cal N}$ is given by,
\beq
\label{mixing}
{\cal N}^*  =  \left(\begin{array}{cc}
N^* & 0\\
0& V_\nu^T \end{array}\right)
\left(\begin{array}{cc}
1 -{1 \over 2} \xi^{\dagger} \xi& \xi^{\dagger} \\
-\xi &  1 -{1 \over 2} \xi \xi^\dagger
\end{array}\right) 
\eeq
The second matrix above block-diagonalize 
${\bold M}_N$ approximately to the form 
diag($m_{eff},{\cal M}_{\chi^0}$)
\beq
m_{eff} = - m \cdot {\cal M}_{\chi^0}^{-1} m^T = 
\frac{M_1 g^2 + M_2 {g'}^2}{4\, det({\cal M}_{\chi^0})} 
\left(\begin{array}{ccc}
\Lambda_e^2 & \Lambda_e \Lambda_\mu
& \Lambda_e \Lambda_\tau \\
\Lambda_e \Lambda_\mu & \Lambda_\mu^2
& \Lambda_\mu \Lambda_\tau \\
\Lambda_e \Lambda_\tau & \Lambda_\mu \Lambda_\tau & \Lambda_\tau^2
\end{array}\right).
\eeq
The submatrices $N$ and $V_{\nu}$ in \Eq{mixing}  diagonalize 
${\cal M}_{\chi^0}$ and $m_{eff}$ 
\beq
N^{*}{\cal M}_{\chi^0} N^{\dagger} = diag(m_{\chi^0_i})
\quad ; \quad
V_{\nu}^T m_{eff} V_{\nu} = diag(0,0,m_{\nu}),
\eeq
where 
\beq
m_{\nu} = Tr(m_{eff}) = 
\frac{M_1 g^2 + M_2 {g'}^2}{4\, det({\cal M}_{\chi^0})} 
|{\vec \Lambda}|^2.
\eeq

\section{One Loop Neutrino Masses and Mixings}
\subsection{Definition}

The Self--Energy for the neutralino/neutrino is

\beq
\hskip 3.5cm \equiv
i \left\{ \slash{p} \left[ P_L \Sigma^L_{ij} + P_R \Sigma^R_{ij} \right]
-\left[ P_L \Pi^L_{ij} + P_R \Pi^R_{ij} \right]\right\}
\eeq

\begin{picture}(0,0)
\put(-0.5,-0.25){\includegraphics[width=3.5cm]{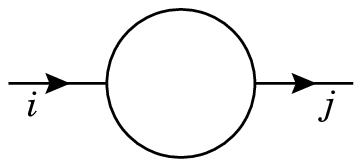}}
\end{picture}

\ni
Then 
\beq
M^{\rm pole}_{ij}= M^{\rm \overline{DR}}_{ij}(\mu_R) + \Delta M_{ij}
\eeq
with
\beq
\Delta M_{ij}\! =\! \left[ \half 
\left(\Pi^V_{ij}(m_i^2)\! +\! \Pi^V_{ij}(m_j^2)\right) 
\!-\! \half 
\left( m_{\chi^0_i} \Sigma^V_{ij}(m_i^2) \! +\! 
m_{\chi^0_j} \Sigma^V_{ij}(m_j^2) \right) \right]_{\Delta=0}
\eeq
where
\beq
\Sigma^V=\half \left(\Sigma^L+\Sigma^R\right)
\quad ; \quad
\Pi^V=\half \left(\Pi^L+\Pi^R\right)
\eeq
and
\beq
\ds \Delta=\frac{2}{4-d} -\gamma_E + \ln 4\pi
\eeq

\subsection{Diagrams Contributing}

In a generic way the diagrams contributing are

\begin{tabular}{cccc}
\begin{picture}(0,3)
\put(-0.2,0){\includegraphics[width=30mm]{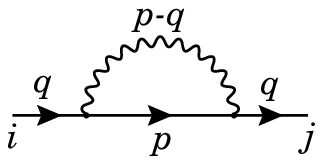}}
\end{picture}&
\begin{picture}(0,0)
\put(2.65,0){\includegraphics[width=30mm]{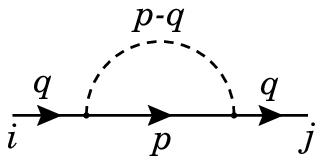}}
\end{picture}&
\begin{picture}(0,0)
\put(5.5,0){\includegraphics[width=30mm]{sself.eps}}
\end{picture}&
\begin{picture}(0,0)
\put(8.35,0.2){\includegraphics[width=30mm]{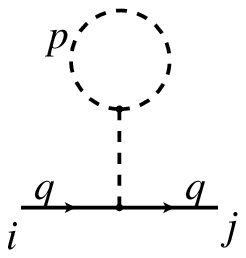}}
\end{picture}
\end{tabular}

\ni
These diagrams can be calculated in a straightforward way. For
instance the $W$ diagram in the $\xi=1$ gauge gives
\beqa
\Sigma^V_{ij}&\!=\!\!& -\frac{1}{16\pi^2}\, \sum_{k=1}^5
2 \left(O^{\rm ncw}_{L jk} O^{\rm cnw}_{L ki} +
O^{\rm ncw}_{R jk} O^{\rm cnw}_{R ki}\right) B_1(p^2,m^2_k,m^2_W)\cr
\vb{35}
\Pi^V_{ij}&\!=\!\!& -\frac{1}{16\pi^2}\, \sum_{k=1}^5
(-4) \left(O^{\rm ncw}_{L jk} O^{\rm cnw}_{R ki} +
O^{\rm ncw}_{R jk} O^{\rm cnw}_{L ki}\right) m_k\, B_0(p^2,m^2_k,m^2_W)
\nn
\eeqa
where $B_0$ and $B_1$ are the Passarino-Veltman functions, and
$O^{\rm cnw}$, $O^{\rm ncw}$ are coupling matrices. Explicit
expressions can be found in~\cite{numass}.

\ignore{
\begin{picture}(0,3.5)
\put(0,0){\includegraphics[height=30mm]{verncw.eps}}
\end{picture}
\vspace{-30mm}

$$
\gamma^{\mu} \left( O^{\rm ncw}_{L ji} P_L +
O^{\rm ncw}_{R ji} P_R \right)
$$

\begin{picture}(0,3.5)
\put(0,0){\includegraphics[height=30mm]{vercnw.eps}}
\end{picture}

\vspace{-30mm}

$$
\gamma^{\mu} \left( O^{\rm cnw}_{L ji} P_L +
O^{\rm cnw}_{R ji} P_R \right)
$$

\vspace{10mm}

}

\subsection{Gauge Invariance}

When calculating the self--energies the question of gauge invariance
arises. We have performed a careful calculation in an arbitrary
$R_{\xi}$ gauge and showed~\cite{numass} that the result was independent of the
gauge parameter $\xi$.

\ignore{
In the $R_{\xi}$ gauge the following diagrams
depend on $\xi$

\begin{tabular}{ccc}
\begin{picture}(0,3)
\put(0,0){\includegraphics[width=30mm]{wself.eps}}
\end{picture}
&
\begin{picture}(0,0)
\put(4,0){\includegraphics[width=30mm]{sself.eps}}
\end{picture}
&
\begin{picture}(0,0)
\put(8,0.2){\includegraphics[width=30mm]{tadgoldstone.eps}}
\end{picture}
\end{tabular}
\begin{center}
\vspace{-5mm}
Set 1
\end{center}

\vspace{5mm}

\begin{tabular}{cc}
\begin{picture}(0,3)
\put(2,0){\includegraphics[width=30mm]{tadzij.eps}}
\end{picture}
&
\begin{picture}(0,0)
\put(7,0){\includegraphics[width=30mm]{tadczij.eps}}
\end{picture}
\end{tabular}

\begin{center}
\vspace{-5mm}
Set 2
\end{center}

\vspace{5mm}

\begin{tabular}{ccc}
\begin{picture}(0,3)
\put(1.25,0){\includegraphics[width=30mm]{tadw.eps}}
\end{picture}
&
\begin{picture}(0,0)
\put(4.25,0){\includegraphics[width=30mm]{tadcw+.eps}}
\end{picture}
&
\begin{picture}(0,0)
\put(7.25,0){\includegraphics[width=30mm]{tadcw-.eps}}
\end{picture}
\end{tabular}
\begin{center}
\vspace{-5mm}
Set 3
\end{center}

\vspace{5mm}

\ni
Calculating in an arbitrary $R_{\xi}$ gauge we have shown~\cite{numass} 
that the gauge dependence cancels among the diagrams in each set.

}

\subsection{The One--Loop Mass Matrix}

The one--loop corrected mass matrix is
\beq
M^{1L}= M^{0L}_{diag} + \Delta M^{1L}
\eeq
where
\beq
M^{0L}_{diag}= {\cal N} M_N {\cal N}^T
\eeq
Now we diagonalize the 1--loop mass matrix 
\beq
M^{1L}_{diag}={\cal N'} M^{1L} {\cal N'}^T
\eeq
Then the mass eigenstates are related to the weak basis states by
\beq
\chi_0^{mass}= {\cal N}^{1L}_{i \alpha}\, \chi_0^{weak}
\eeq
with
\beq
{\cal N}^{1L} = {\cal N'}\ {\cal N}
\eeq

\ni
The usual convention in neutrino physics
\beq
\nu_{\alpha} = U_{\alpha k}\,  \nu_k
\eeq
is recovered in our notation as
\beq
U_{\alpha k}= {\cal N}^{1L}_{4+k, 4+\alpha} 
\eeq

\subsection{Solar and Atmospheric Neutrino Parameters}

Assuming hierarchy in the masses $m_{\nu_2}$ and $m_{\nu_3}$
and neglecting $U_{e3}$ (that has to be small) 
we write the usual two neutrino mixing angle as
\ignore{
the
survival probabilities for the solar and atmospheric neutrinos 
are
\begin{eqnarray}
P_e&=&1 - 4 U_{e1}^2 U_{e2}^2\ \sin^2 \left( \frac{\Delta m^2_{21}t}{4 E}
\right)
-2 U_{e3}^2 (1 - U_{e3}^2 )\nn \\
P_{\mu}&=&1 - 4 U_{\mu 3}^2 (1 -U_{\mu 3}^2)\ 
\sin^2 \left( \frac{\Delta m^2_{21}t}{4 E}
\right)
\end{eqnarray}
As $U_{e3}$ has to be small we neglect it and write the usual two
neutrino mixing angle as
}
\beq
\sin^2 (2 \theta_{12}) = 4\, U_{e1}^2 U_{e2}^2
\eeq
and
\beq
\sin^2 (2 \theta_{13}) = 4\, U_{\mu 3}^2 (1 -U_{\mu 3}^2)
\eeq

\section{Results}

\subsection{Masses and Mixings}

\begin{figure}[htb]
\begin{center}
\begin{picture}(9,6)
\put(0,0){\includegraphics[width=9cm]{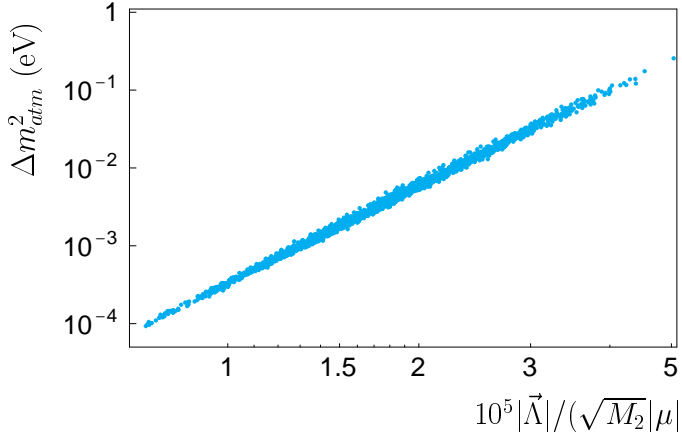}}
\end{picture}
\end{center}
\vspace{-5mm}
\caption{The atmospheric $\Delta m^2$ as function of 
$|\vec \Lambda |/(\sqrt{M_2}\mu)$}
\label{cascais2000fig3}
\end{figure}

\begin{figure}[htb]
\begin{center}
\begin{picture}(9,6)
\put(0,0){\includegraphics[width=9cm]{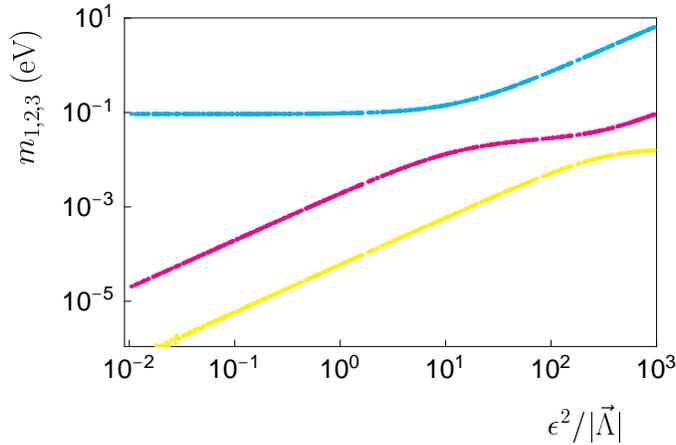}}
\end{picture}
\end{center}
\vspace{-5mm}
\caption{Neutrino masses as a function of $\epsilon^2/|\vec \Lambda|$}
\label{allmasses}
\end{figure}

The BRPV model produces a hierarchical mass spectrum for almost all
choices of parameters. The largest mass can be estimated by the tree
level value as shown in \Fig{cascais2000fig3}. As the figure shows,
correct $\Delta m^2_{atm}$ can be easily obtained by an appropriate
choice of $| \vec \Lambda|$. The mass scale for the solar neutrinos is
generated a 1--loop level and therefore depends in a complicated way
in the model parameters. This is shown in \Fig{allmasses} where
we have fixed the SUSY parameters. The parameter $\epsilon^2/|\vec \Lambda|$ 
is the most important in determining the solar mass scale, but some
other parameters also play a role~\cite{numass}.

\begin{figure}[htb]
\begin{tabular}{cc}
\begin{picture}(5.825,5.5)
\put(0,0){\includegraphics[width=5.9cm,height=55mm]{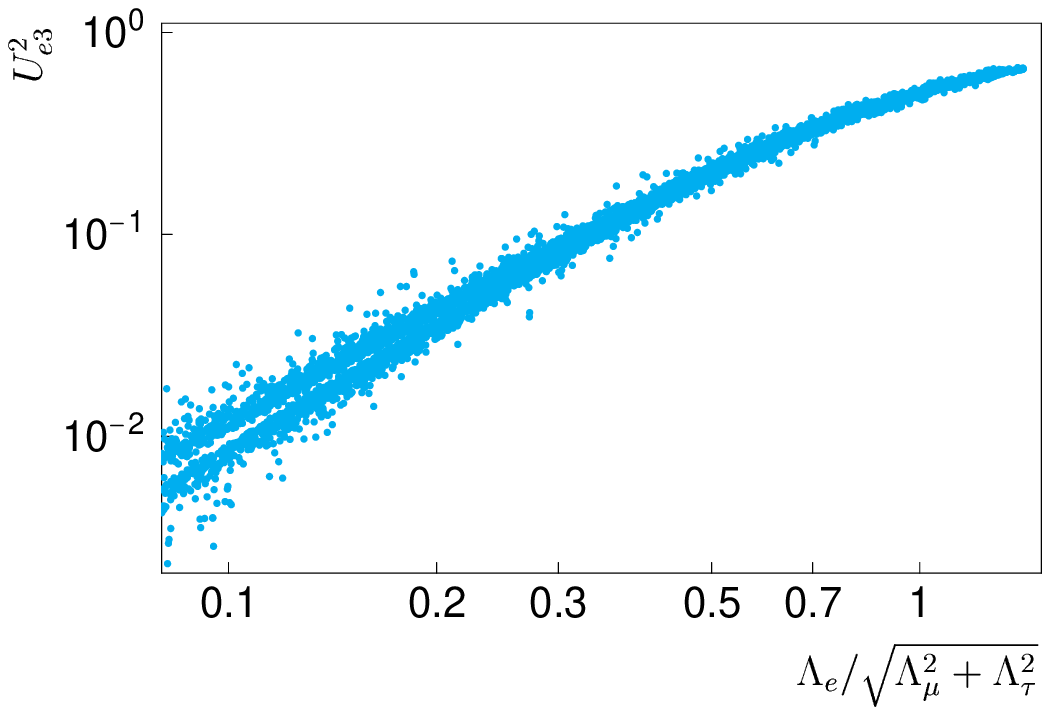}}
\end{picture}
&
\begin{picture}(5.825,5.5)
\put(0,0){\includegraphics[width=5.9cm,height=55mm]{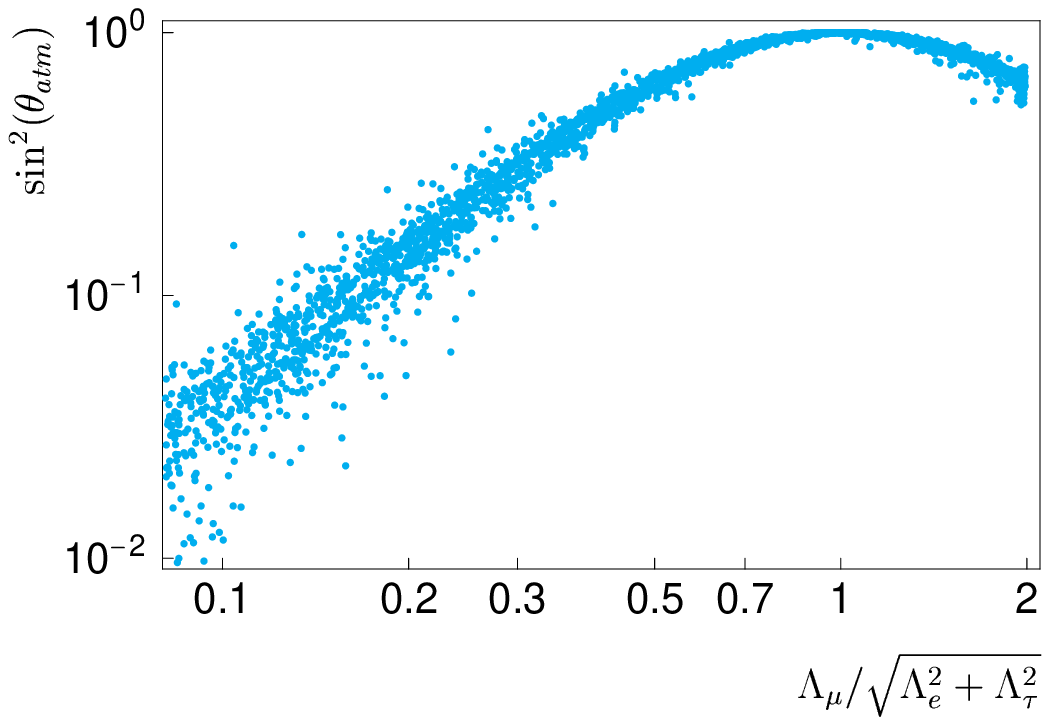}}
\end{picture}
\end{tabular}
\vspace{-5mm}
\caption{ a) $U^2_{e3}$ as a function of 
$|\Lambda_e|/\sqrt{\Lambda^2_{\mu}+\Lambda^2_{\tau}}$.
b) Atmospheric angle as a function of 
$|\Lambda_{\mu}|/\sqrt{\Lambda^2_{\mu}+\Lambda^2_{\tau}}$.}
\label{val2000fig56}
\end{figure}

\ni
Now we turn to the discussion of the mixing angles. We have found that if 
$\epsilon^2/|\vec \Lambda| \ll 100$ then the 1--loop corrections are
not larger then the tree level results and the flavour composition of
the 3rd mass eigenstate is approximately given by
\beq
U_{\alpha 3}\approx\Lambda_{\alpha}/|\vec \Lambda |
\eeq
As the atmospheric and reactor neutrino data tell us that
$\nu_{\mu}\ra \nu_{\tau}$ oscillations are preferred over 
$\nu_{\mu}\ra \nu_e$, we conclude that  
\beq
\Lambda_e \ll \Lambda_{\mu} \simeq \Lambda_{\tau}
\eeq
are required for BRPV to fit the the data. This is sown in
\Fig{val2000fig56}.

\begin{figure}[htb]
\begin{tabular}{cc}
\begin{picture}(5.825,5)
\put(0,0){\includegraphics[width=5.825cm,height=50mm]{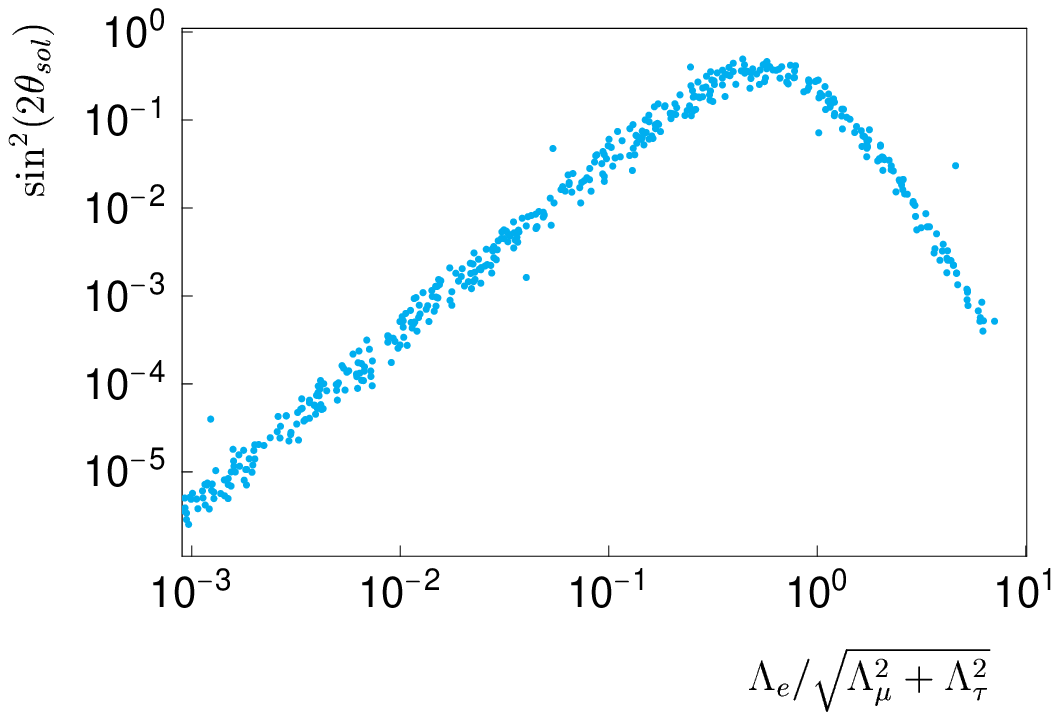}}
\end{picture}
&
\begin{picture}(5.825,5)
\put(0,0){\includegraphics[width=5.825cm,height=50mm]{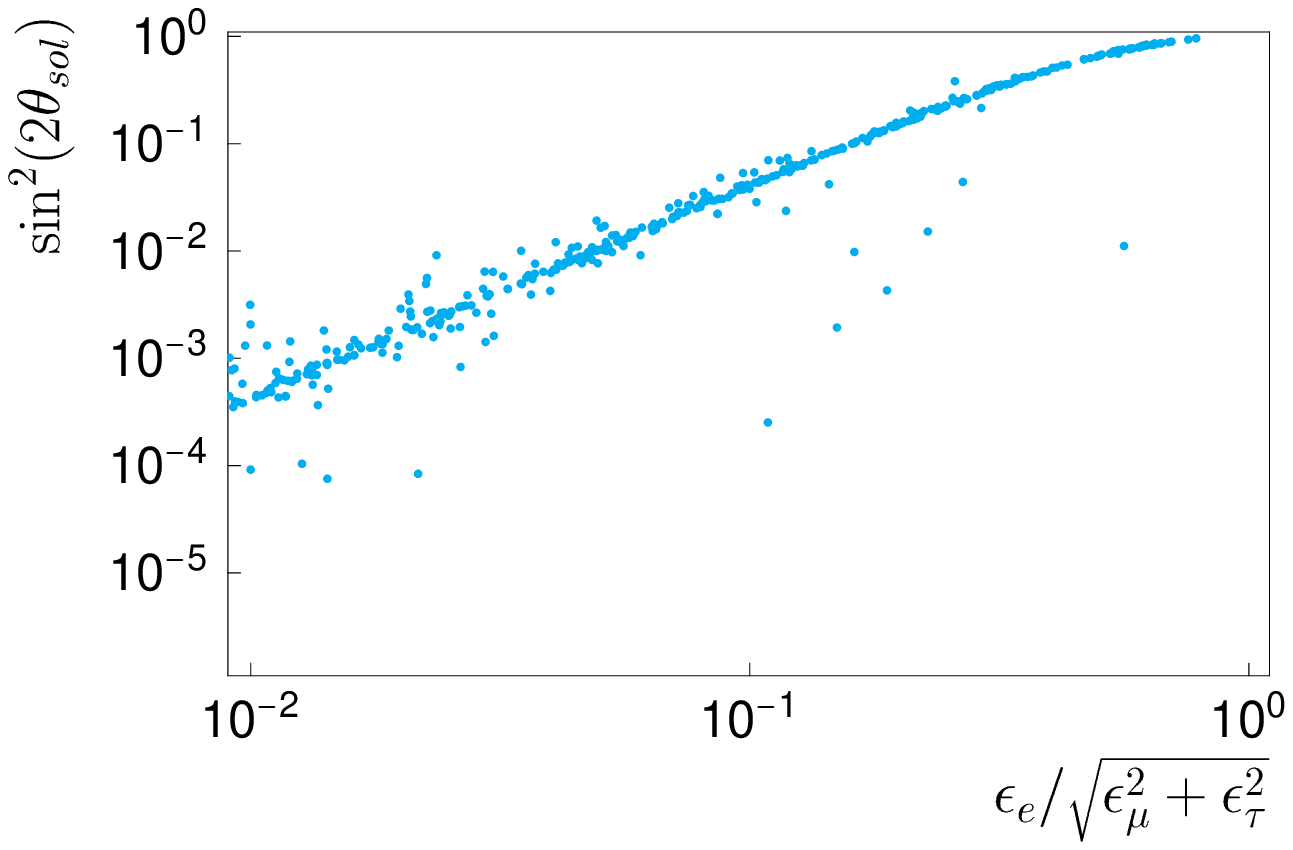}}
\end{picture}
\end{tabular}
\vspace{-5mm}
\caption{The solar angle as function of:
a)$|\Lambda_e|/\sqrt{\Lambda^2_{mu}+\Lambda^2_{\tau}}$;
b)$\epsilon_e/\sqrt{\epsilon^2_{mu}+\epsilon^2_{\tau}}$}
\label{val2000fig78}
\end{figure}

\ni 
For the solar angle the situation is more complicated and there are two
cases to consider~\cite{numass}. With the usual SUGRA assumptions,
ratios of $\epsilon_i/\epsilon_j$ fix the ratios of $\Lambda_i/\Lambda_j$.
Since atmospheric and reactor data tell us that $\Lambda_e \ll
\Lambda_{\mu},\Lambda_{\tau}$ in this case only the small angle
solution can be obtained in the BRPV model as shown in
Fig.~(\ref{val2000fig78} a).
However we have shown that
even a tiny deviation of from universality of the soft parameters at
the GUT scale relaxes this constraint. In this case the ratio
$\epsilon_i/\epsilon_j$ is not constrained and 
also large angle
solutions can be obtained as shown in Fig.~(\ref{val2000fig78} b).

\subsection{Consequences for the accelerators}

\begin{figure}[htb]
\begin{tabular}{cc}
\begin{picture}(5.8,5.5)
\put(0,0){\includegraphics[width=5.8cm,height=5.5cm]{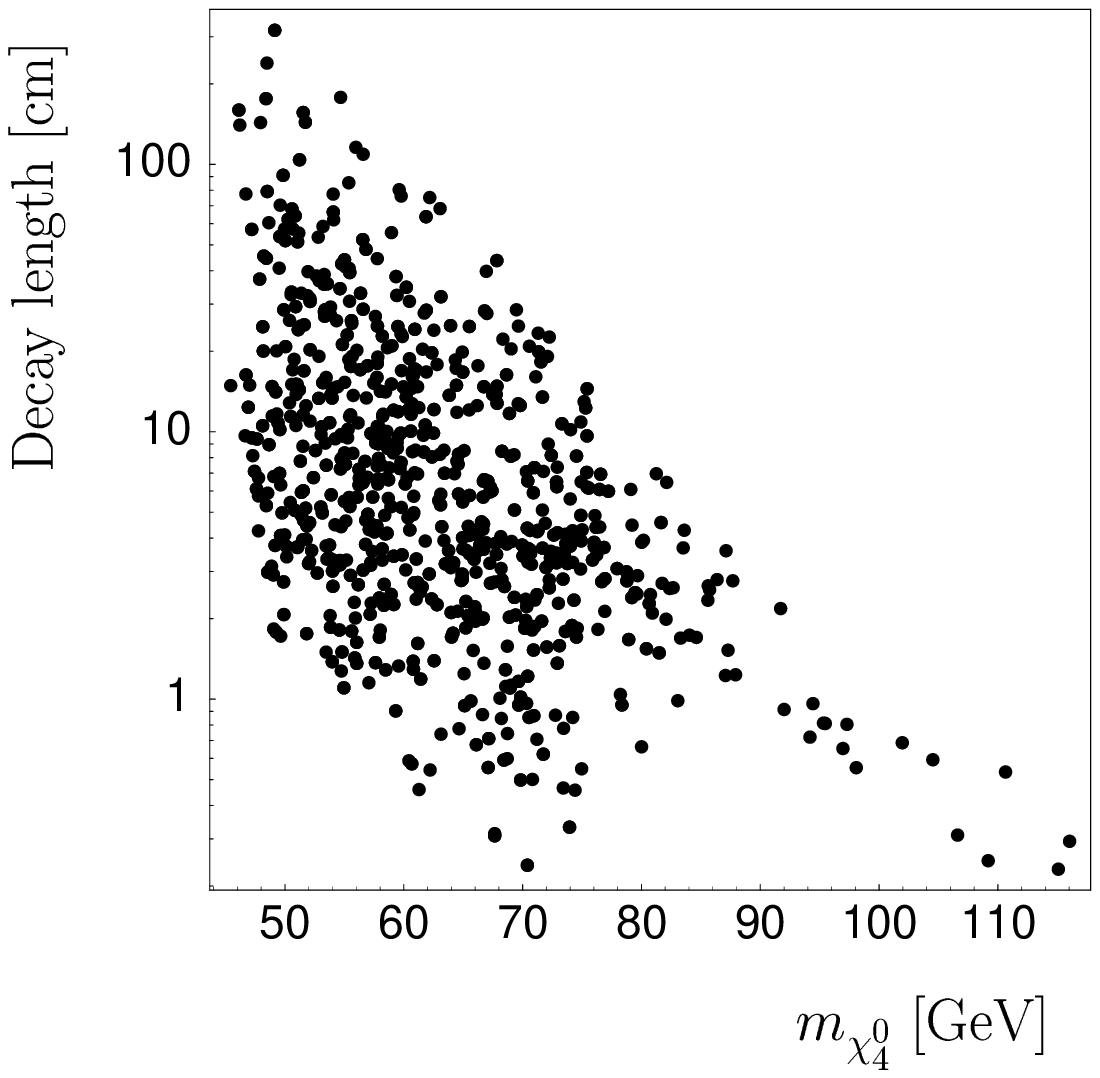}}
\end{picture}
&
\begin{picture}(5.8,5.5)
\put(0,0){\includegraphics[width=5.8cm,height=5.5cm]{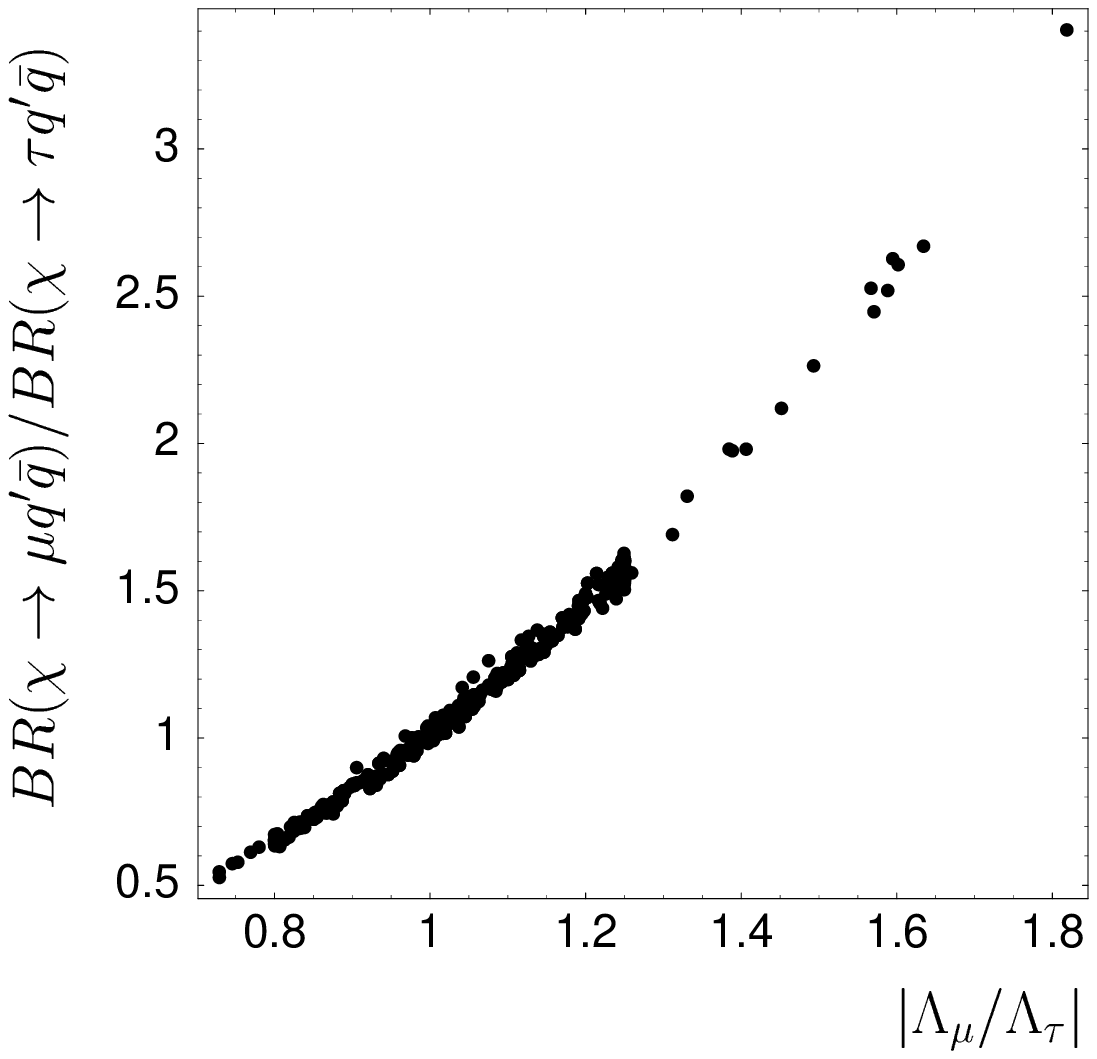}}
\end{picture}
\end{tabular}
\vspace{-3mm}
\caption{a) Neutralino decay length; 
b) $BR(\chi \to \mu q' \bar q)/ BR(\chi \to \tau q' \bar q$) 
as function of $\Lambda_\mu/\Lambda_\tau$}
\label{pisa_fig12}
\end{figure}

One of the attractives of the BRPV model is that besides accommodating
the solar and atmospheric neutrino data it can make definite predictions
for accelerator physics. Then the model can be tested.
As R-parity is violated, the neutralino is unstable. For this to have
experimental consequences the neutralino has to decay well inside the
detector. This is indeed the case as shown in 
Fig.~(\ref{pisa_fig12} a). 
\begin{figure}[htb]
\begin{tabular}{cc}
\begin{picture}(5.8,5.5)
\put(0,0){\includegraphics[width=5.8cm,height=5.5cm]{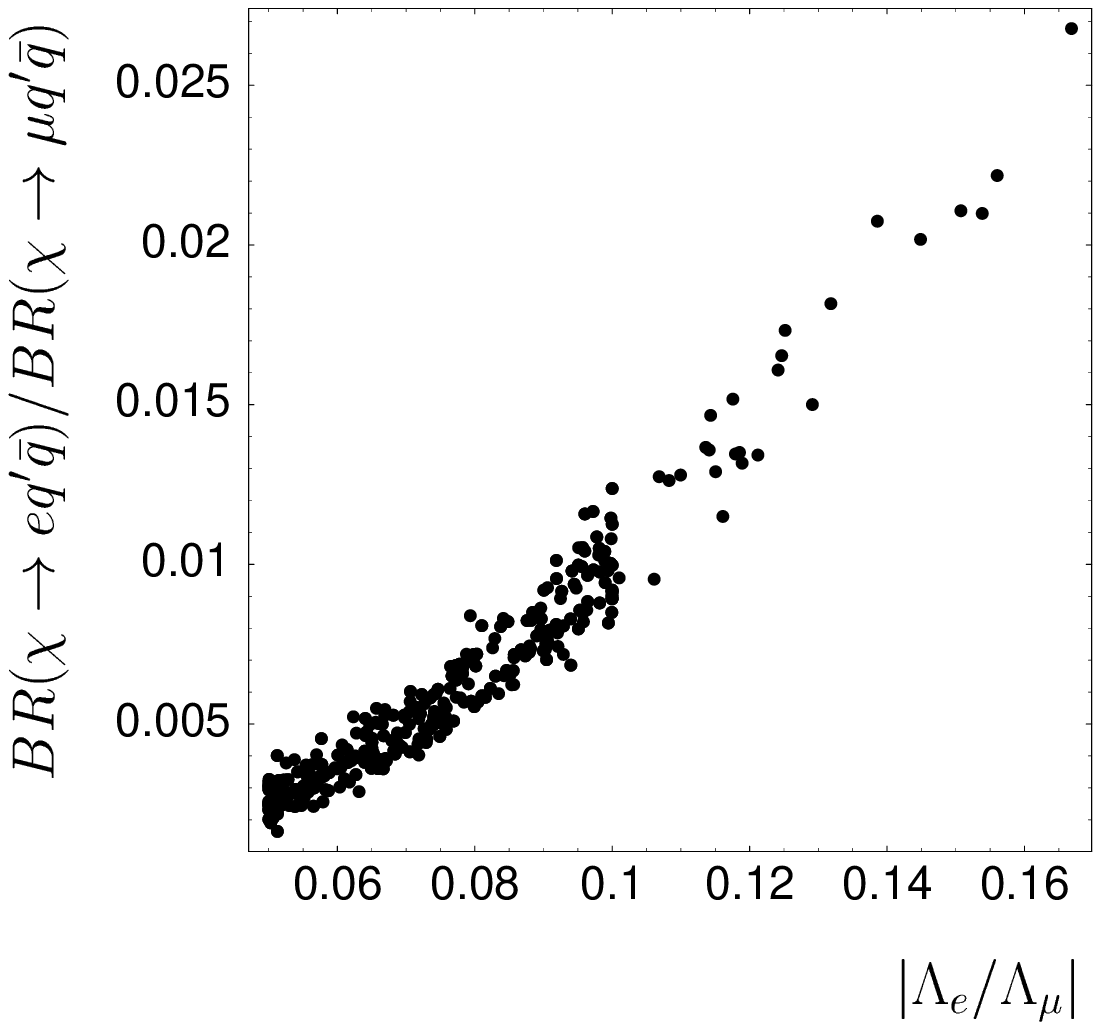}}
\end{picture}
&
\begin{picture}(5.8,5.5)
\put(0,0){\includegraphics[width=5.8cm,height=5.5cm]{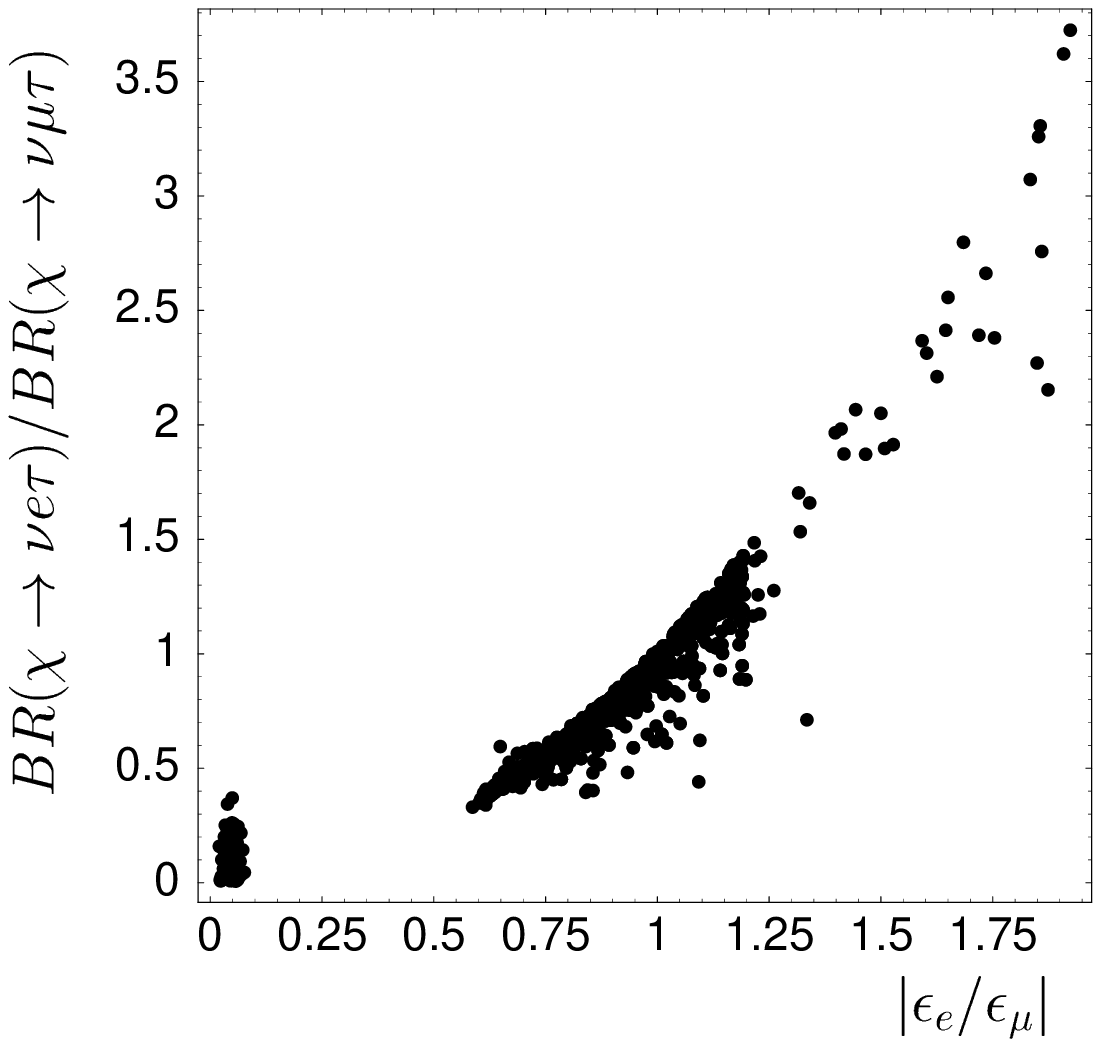}}
\end{picture}
\end{tabular}
\vspace{-3mm}
\caption{
Ratio of branching ratios for semileptonic (a) and leptonic (b) LSP
decays into $\mu$ and $\tau$ as functions of the BRPV parameters.}
\label{pisa_fig34}
\end{figure}
We have seen before that the ratios
$|\Lambda_i/\Lambda_j|$ and $\epsilon_i/epsilon_j|$ were very
important in the choice of solutions for the neutrino mixing
angles. What is exciting is that this ratios can be measured in
accelerator experiments. 
In Fig.~(\ref{pisa_fig12} b) we show the 
ratio of branching ratios for semileptonic LSP
decays into muons and taus: $BR(\chi \to \mu q' \bar q)/ BR(\chi \to
\tau q' \bar q$) as function of $\Lambda_\mu/\Lambda_\tau$. We can see
that this is directly correlated with atmospheric angle.
In Fig.~(\ref{pisa_fig34} a) is shown the
ratio of branching ratios for semileptonic LSP
decays into muons and taus: $BR(\chi \to e q' \bar q)/ BR(\chi \to
\mu q' \bar q$) as function of $\Lambda_e/\Lambda_\mu$. This is
directly correlated with $U_{e3}^2$. 
Finally in Fig.~(\ref{pisa_fig34} b) we show the
ratio of branching ratios for leptonic LSP
decays into muons and taus: 
$BR(\chi \to \nu e \tau)/ BR(\chi \to
\nu \mu \tau)$ 
as function of $|\epsilon_{e}/\epsilon_{\mu}|$. This is
directly correlated with solar angle.

\section{Conclusions}

We have shown that there is
a viable model for SBRP that leads to a very rich
phenomenology, both at laboratory experiments, and at pre\-sent (LEP)
and future (LHC, NLC) accelerators.
Most of these phenomenology can be described by an effective model
with explicit R--Parity violation.
We have calculated the one--loop corrected masses
and mixings for the
neutrinos in a completely consistent way, including the RG equations
and correctly minimizing the potential.
We have shown that it is possible to get easily maximal mixing for the
atmospheric neutrinos and both small and large angle MSW.
We emphasise that the 
lightest neutralino decays inside the detectors,
thus leading to a very different phenomenology than the MSSM. 
If the model is to explain solar and atmospheric neutrino problems
then many signals will arise at future colliders. 
Thus the model can easily be proved wrong.


\begin{thebibliography}{}  

\bibitem{Fukuda:1998mi}
Fukuda, Y. {\it et al.}  [Super-Kamiokande Collaboration] (1998),
{\it Phys. Rev. Lett.} {\bf 81}, 1562, hep-ex/9807003; see also
hep-ex/9803006 and hep-ex/9805006.

\bibitem{Gonzalez-Garcia:1998hj}
Gonzalez-Garcia, M.C., Nunokawa H., Peres, O.L. and
Valle, J.W. (1999) {\it Nucl. Phys.} {\bf B543}, 3; Gonzalez-Garcia, M.C.
{\it et al.} (1999), {\it Phys. Rev. Lett.} {\bf 82}, 3202, hep-ph/9809531;
Fornengo, N., Gonzalez-Garcia, M.C. and Valle, J.~W.~F. (1999), hep-ph/9906539;
Barger, V. Learned, J.G., Pakvasa, S., and Weiler, T.J. (1999),
{\it Phys. Rev. Lett.} {\bf 82}, 2640.
 
\bibitem{BP98} 
Bahcall, J. N., Basu  S. and Pinsonneault, M. H. (1998), {\it Phys. Lett.} 
{\bf B 433}, 1.

\bibitem{Smy:1999tt}
Smy, M.B. (1999) ``Solar neutrinos with SuperKamiokande," hep-ex/9903034.

\bibitem{MSW99} 
For an updated discussion of solar and atmospheric neutrino data see
the talk of M.C. Gonzalez-Garcia at the ICHEP 2000. An extended
version of the talk can be found in 
Gonzalez-Garcia, M.C. and  Pe\~na-Garay, C. (1999) hep-ph/0009401.

\bibitem{numass}
Rom\~ao, J.C., D\'{\i}az, M.A., Hirsch, M., Porod, W. and Valle, J.W.F. (2000)
{\it Phys. Rev.}{\bf D61}, 071703 (2000);
Hirsch, M., D\'{\i}az, M.A., Porod, W., Rom\~ao, J.C. and Valle, J.W.F. (2000),
hep-ph/0004115 to appear in {\it Phys. Rev.}{\bf D}, 


\bibitem{epsmodel}
D\'{\i}az, M.A., Rom\~ao, J.C. and Valle, J.W.F. (1998) {\it Nucl. Phys.}
{\bf 524}, 23.


\bibitem{brasil98}
For a review see {\it e.g} 
Rom\~ao, J. C. (1998) Lectures given at the 
{\it 5th Gleb Wataghin School on High-Energy Phenomenology}, Campinas, Brazil, 13-17 Jul 1998,  hep-ph/9811454.

\bibitem{OriginalSBRP}
Aulakh, C., Mohapatra, R. (1983) {\it Phys. Lett.} {\bf B119}, 136;
Santamaria, A., Valle, J.W.F. (1987) {\it Phys. Lett.} {\bf B195}, 423;
Santamaria, A., Valle, J.W.F. (1988) {\it Phys. Rev. Lett.} {\bf 60}, 397.

\bibitem{SBRpV}
Masiero, A. and Valle, J.W.F. (1990) {\it Phys. Lett.} {\bf B251}, 273;
Rom\~ao, J.C., Santos, C.A. and Valle, J.W.F. (1992) {\it Phys. Lett}
{\bf B288}, 311.


\bibitem{paulo+npb} 
Nogueira, P., Rom\~ao, J.C. and Valle, J.W.F (1990) {\it Phys. Lett} 
{\bf B251}, 142;
Rom\~ao, J.C. and Valle, J.W.F. (1992) {\it Nucl. Phys.} 
{\bf B381}, 87.

\bibitem{bbnothers}
Bertolini, S. and Steigman, G. (1990) {\it Nucl. Phys.} 
{\bf B387}, 193;
Kawasaki, M.  {\it et al} (1993) {\it Nucl. Phys.} 
{\bf B402}, 323;
Kawasaki, M.  {\it et al} (1994) {\it Nucl. Phys.} 
{\bf B419}, 105;
Dodelson, S., Gyuk, G. and Turner, M.S. (1994) {\it Phys. Rev.}
{\bf D49}, 5068.

\bibitem{bbnpaper}
Dolgov, A.D., Pastor, S., Rom\~ao J.C. and Valle, J.W.F. (1997)
{\bf B496}, 24.



\bibitem{marco}
D\'{\i}az, M.A., (1998) Proceedings of the {\it International
Workshop on Physics Beyond the Standard Model: From Theory to
Experiment} (Valencia 97), Eds. Antoniadis, I., Iba\~nez, L.E. and
Valle, J.W.F., World Scientific, pag. 188. 



\end{thebibliography}
\end{document}
\footnote{